\documentclass[twocolumn,superscriptaddress,aps,prl]{revtex4-1}

\usepackage{amsmath}
\usepackage{graphics}
\usepackage[version=3]{mhchem}
\usepackage{siunitx}
\usepackage[caption=false]{subfig}
\usepackage{rotating}
\usepackage{float}
\usepackage{hyperref}
\usepackage{natbib}

\begin{document}
\title{High Energy Density Mixed Polymeric Phase From Carbon Monoxide And
Nitrogen}
\author{Zamaan Raza}
\affiliation{Institut de Min\'{e}ralogie et de Physique des Milieux
Condens\'{e}s (IMPMC), Universit\'{e} Pierre et Marie Curie, 4 place Jussieu,
75252 Paris cedex 05, France}
\author{Chris J.\ Pickard}
\affiliation{Department of Physics and Astronomy, University College London,
London WC1E 6BT, UK}
\author{Carlos Pinilla}
\affiliation{Institut de Min\'{e}ralogie et de Physique des Milieux
Condens\'{e}s (IMPMC), Universit\'{e} Pierre et Marie Curie, 4 place Jussieu,
75252 Paris cedex 05, France}
\author{A.\ Marco Saitta}
\affiliation{Institut de Min\'{e}ralogie et de Physique des Milieux
Condens\'{e}s (IMPMC), Universit\'{e} Pierre et Marie Curie, 4 place Jussieu,
75252 Paris cedex 05, France}
\date{\today}

\begin{abstract}
Carbon monoxide and nitrogen are among the potentially interesting high energy
density materials. However, in spite of the physical similarities of the
molecules, they behave very differently at high pressures. Using density
functional theory and structural prediction methods, we examine the ability of
these systems to combine their respective properties and form novel mixed
crystalline phases under pressures of up to \SI{100}{\giga\pascal}.
Interestingly, we find that \ce{CO} catalyzes the molecular dissociation of
\ce{N_2}, which means mixed structures are favored at a relatively low pressure (below
\SI{18}{\giga\pascal}) and that a three-dimensional framework with $Pbam$
symmetry becomes the most stable phase above \SI{52}{\giga\pascal}, i.e.\@ at
much milder conditions than in pure solid nitrogen. This structure is
dynamically stable at ambient pressure, and has an energy density of
approximately \SI{2.2}{\kilo\joule\per\gram}, making it a candidate for a
high-energy density material, and one that could be achieved at less
prohibitive experimental conditions.
\end{abstract}

\maketitle
\graphicspath{{./Figures/}}

Both carbon monoxide and nitrogen form molecular crystals that irreversibly
polymerize at high pressures to form dense single bonded frameworks with
high-energy densities. It has been shown that \ce{CO} undergoes the transition
from a molecular crystal to a polymeric phase at much lower pressures than
nitrogen\citep{Lipp2005,Eremets2004}. While both high pressure \ce{CO} and N
are potential high-energy density materials (HEDM), polymeric \ce{CO} has been
found to decompose under ambient conditions, and polymeric N can only be formed
at prohibitively high pressures and has not been recovered to ambient
pressure\citep{Eremets2004a}. In this Letter, we present calculations
demonstrating that a mixture of \ce{CO} and \ce{N_2} will undergo a transition
to a van der Waals-bound polymeric phase at modest pressures compared with
nitrogen. This material will undergo a further transition as the pressure is
increased further, to a high-energy density three-dimensional framework, which
is expected to be metastable at ambient pressure.

Our interest in the CNO system is also motivated by recent success in the
experimental synthesis of such condensed molecular
materials\citep{McMillan2002}, some of which are HEDM\citep{HEDM,Lipp2005}.
This class of low-Z materials formed at high pressures is also of interest for
reasons beyond HEDM; they may have other useful properties such as
superhardness in the case of polymeric \ce{CO_2}\citep{Yoo1999,Datchi2012},
superconductivity and optical nonlinearity\citep{Lipp2005}.  They are crucial
in our understanding of high-pressure carbon chemistry, which in turn is vital
for projects such as the Deep Carbon Observatory (DCO), concerned with carbon
capture and sequestration. Bonds involving carbon, nitrogen and oxygen form the
backbone of organic biochemistry, and this work will help improve our
understanding of their chemistry.

\ce{CO} and \ce{N_2} molecules are very similar. \ce{CO} has the
strongest known chemical bond and \ce{N_2} has the strongest homonuclear
bond\citep{Sun2011}. They are isoelectronic, have the same total mass number,
have moments of inertia that are within 3\%\citep{Witters1977}, and have
qualitatively similar phase diagrams\citep{Mills1986}. Although they are
similar at low and ambient pressures, differences in chemical bonding leads to
a divergence at high pressures, a regime in which \ce{CO} remains poorly
experimentally characterized. They also differ in symmetry and charge
distribution; CO has a small dipole moment, but its quadrupole moment is
significant, with consequences in its phase diagram\citep{Serdyukov2010}. For
example, quadrupole considerations forbid the formation of a CO phase analogous
to $\gamma$-\ce{N_2}\citep{Felsteiner1971}.

In spite of these similarities, the literature contains only a small amount of
work on \ce{CO}/\ce{N_2} mixtures, limited to Raman studies on low temperature
and ambient pressure \ce{CO}/\ce{N_2} and experimental work on orientation and
substitutional disorder in molecular \ce{CO}/\ce{N_2} crystals\citep{Turc1992}.
Ab initio molecular dynamics simulations of a liquid \ce{CO}/\ce{N_2} mixture
suggest that molecular dissociate happens at much lower temperatures and
pressures (\SI{3500}{\kelvin}, \SI{20}{\giga\pascal}) than for pure liquid
nitrogen\citep{Chen2012}.

At low temperatures and pressures, nitrogen forms molecular crystals weakly
bound by van der Waals interactions. As the pressure is increased, the
molecules dissociate, resulting in single-bonded covalent solids with
three-coordinated atoms. The cubic gauche (cg-N) nonmolecular phase was
predicted by Mailhiot \textit{et al}.\@ via distortions of a simple cubic
lattice\citep{Mailhiot1992}, and first synthesized over ten years later in
diamond anvil cell experiments at high-pressure and temperature
(\SI{110}{\giga\pascal}, \SI{2000}{\kelvin})\citep{Eremets2004}. There is
striking disagreement in the stability regime of cg-N between theory
(\SI{50}{\giga\pascal}\citep{Mailhiot1992} to under
\SI{100}{\giga\pascal}\citep{McMahan1985}) and experiment (above
\SI{94}{\giga\pascal}\citep{Eremets2001}); it has been suggested that this is a
result of approximate density functional theory (DFT), incorrect nitrogen
structures in calculations and high kinetic barriers to molecular
dissociation\citep{Pickard2009}. Indeed, phonon calculations indicate the cg-N
structure is dynamically stable across a large pressure range, and that there
is a \SI{0.86}{\electronvolt} barrier separating cg-N from the $\beta$-\ce{O_2}
phase of nitrogen at ambient pressure (compared with approximately
\SI{0.3}{\electronvolt} between diamond and graphite)\citep{Barbee1993}.

There is a great deal of interest in high pressure nitrogen phases due to the
large difference between the energy of a single bond and a third of the energy
of a triple bond. This is predicted to result in a large energy release when
cg-N decomposes to molecular nitrogen at ambient pressure, corresponding to an
energy density of approximately \SI{9.7}{\kilo\joule\per\gram}\citep{Uddin2006}
(including a zero-point energy correction), around three times higher than the
most powerful conventional explosives\citep{Zahariev2006}. Its decomposition to
harmless and undetectable nitrogen gas makes it is an excellent candidate for a
rocket propellant; however, DFT predicts that it is metastable at ambient
pressure, while experiments have only succeeded in recovering it to around
\SI{50}{\giga\pascal} to date\citep{Eremets2004a}. Recent MP2 and dispersion
corrected hybrid DFT calculations indicate that the domain of thermodynamic
stability extends only as low as \SI{62}{\giga\pascal} at low
temperatures\citep{Erba2011}.

Carbon monoxide transforms to a polymerized, non-crystalline form at
\SI{5}{\giga\pascal} and \SI{300}{\kelvin}\citep{Lipp2005}, undemanding
conditions compared with nitrogen. It has since been suggested that this
polymeric form of \ce{CO} is in fact not a high-pressure phase, even though it
was synthesized at pressure\citep{Brazhkin2006}. The positions of the stability
lines on the \ce{CO} phase diagram are at much lower pressures and temperatures
compared with other simple molecular solids such as \ce{N_2}, \ce{N_2O} and
\ce{CO_2}\citep{Serdyukov2010}. In fact, CO is predicted to polymerize to
polycarbonyl chains even at ambient pressure and low
temperatures\citep{Sun2011}. Polymeric CO decomposes to molecular \ce{CO_2} and
graphitic carbon, releasing a large amount of energy, in excess of the
explosive HMX. It has been recovered to ambient conditions, at which point it
is metastable\citep{Evans2006}.

Thus far, only amorphous CO has been synthesized in experiments at high
pressure; however nonmolecular crystalline phases have been predicted, in
particular, the $I2_12_12_1$ single-bonded three-dimensional framework is
stable in the \SIrange{2}{55}{\giga\pascal} range, and the layered $Cmcm$
configuration is favored above \SI{55}{\giga\pascal}\citep{Sun2011}.  The
$Cmcm$ structure is particularly interesting because the interlayer separation
and band gap can be modified by varying the pressure, such that it becomes a
semiconductor with a \SI{0.1}{\electronvolt} band gap at approximately
\SI{150}{\giga\pascal}\citep{Sun2011}.

We used the AIRSS method to find candidates for the most stable structures
containing carbon monoxide and nitrogen molecules. AIRSS employs random
structure searching by generating a large ensemble of random ``sensible
structures,'' which may involve the application of biases such as symmetry,
structural units or experimentally derived lattice
parameters\citep{Pickard2006}. At the most basic level, we construct unit cells
with random lattice vectors and place atoms in the chosen stoichiometry at
random positions. This lattice vectors and atomic positions of these cells are
fully relaxed to the local minimum using DFT, and the enthalpies compared to in
principle find the global minimum. The method is described in detail in
reference \citep{Pickard2011}. It has been applied successfully in the
determination of, among others, high-pressure structures of silane
\citep{Pickard2006}, nitrogen\citep{Pickard2009}, carbon
monoxide\citep{Sun2011} and recently a unique
cagelike diamondoid nitrogen phase\citep{Wang2012}.

In the first instance, we generated at least 1000 structures for 1, 2, 3 and 4
formula units of \ce{CON_2} at pressures of 20, 60 and \SI{100}{\giga\pascal},
since these pressures are easily experimentally accessible and are most likely
to yield high energy density materials in the form of a three-dimensional
framework.  Additional searches with larger unit cells were biased towards
high-symmetry structures by construction using either two or four symmetry
operations.  Later, structures were generated using 1--4 formula units of
\ce{2(CO)N_2}.  All structures were optimized using the Perdew-Burke-Ernzerhof
(PBE) exchange correlation functional; after ranking by enthalpy, the best
structures were reoptimized at a higher precision\footnote{All structures were
relaxed at the generalised gradient approximation (GGA) level using the
Perdew-Burke-Ernzerhof (PBE) exchange correlation functional.  Valence
electrons were represented in a plane wave basis set with a cutoff of
\SI{340}{\electronvolt}, and core electrons were modelled using ultrasoft
pseudopotentials which were tested against hard pseudopotetials comparable to
all-electron quality. Monkhorst-Pack k-point grids were generated on a per cell
basis, using a spacing of 2$\pi\times$\SI{0.07}{\per\angstrom}. All DFT
calculations were performed using the CASTEP code\citep{Clark2005}.  Higher
precision calculations were performed using a plane wave cutoff of
\SI{500}{\electronvolt}, a k-point grid spacing of
2$\pi\times$\SI{0.03}{\per\angstrom} and tighter convergence tolerances. Phonon
calculations were performed using density functional perturbation theory (DFPT)
as implemented in CASTEP\citep{Refson2006}, on a $3\times5\times9$
Monkhorst-Pack grid, using norm-conserving pseudopotentials and a plane wave
cutoff of \SI{900}{\electronvolt}, in order to test the dynamical stability of
the very best structures. In order to avoid a core overlap resulting from the
short length of the N$_2$ bond, decomposition enthalpies were computed using a
harder pseudopotential.}.

We first note that the appearance of a mixed carbon monoxide and nitrogen phase
occurs at low pressures compared with pure nitrogen, which forms molecular
crystals up to \SI{56}{\giga\pascal}. Carbon monoxide is a much more reactive
molecule, and appears to facilitate polymerizations even at low pressures.

There are two main pressure regimes of interest below \SI{100}{\giga\pascal}.
At approximately \SI{20}{\giga\pascal} and below, for both stoichiometries,
one-dimensional planar polymers based on an unsaturated carbon/nitrogen
six-ring are favored. At pressures above \SI{20}{\giga\pascal}, fully covalent
three-dimensional frameworks are preferred. In a 1$:$1 mixture of \ce{CO} and
\ce{N_2}, the aforementioned polymers are intermixed with nitrogen molecules,
suggesting a degree of phase separation that can be removed by reducing the
proportion of nitrogen.

\begin{figure}[hb]
  \subfloat[\SI{20}{\giga\pascal}\label{fig:hull-20GPa}]{\begin{turn}{-90}\includegraphics[width=0.20\textwidth]{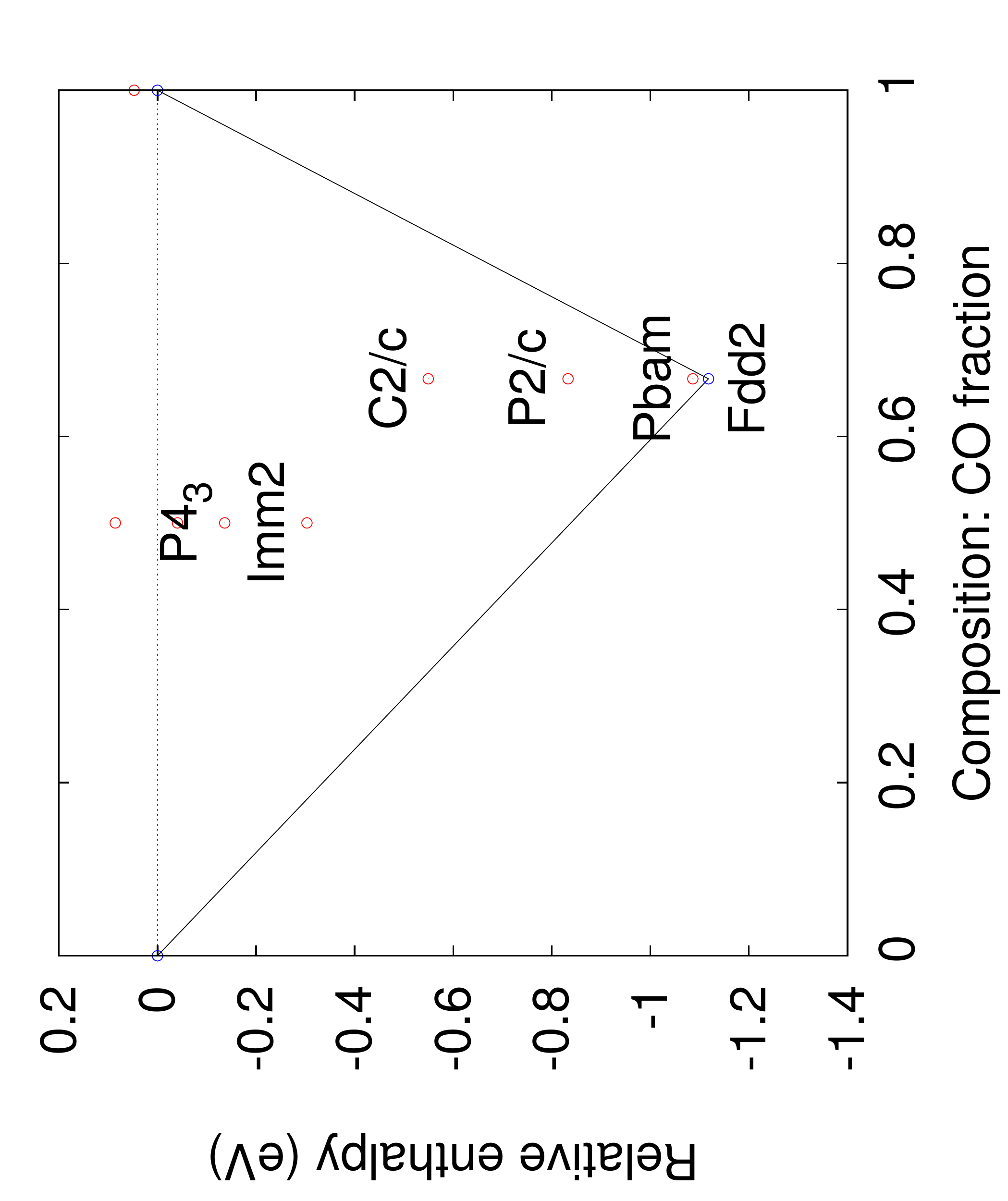}\end{turn}}
  \subfloat[\SI{100}{\giga\pascal}\label{fig:hull-100GPa}]{\begin{turn}{-90}\includegraphics[width=0.20\textwidth]{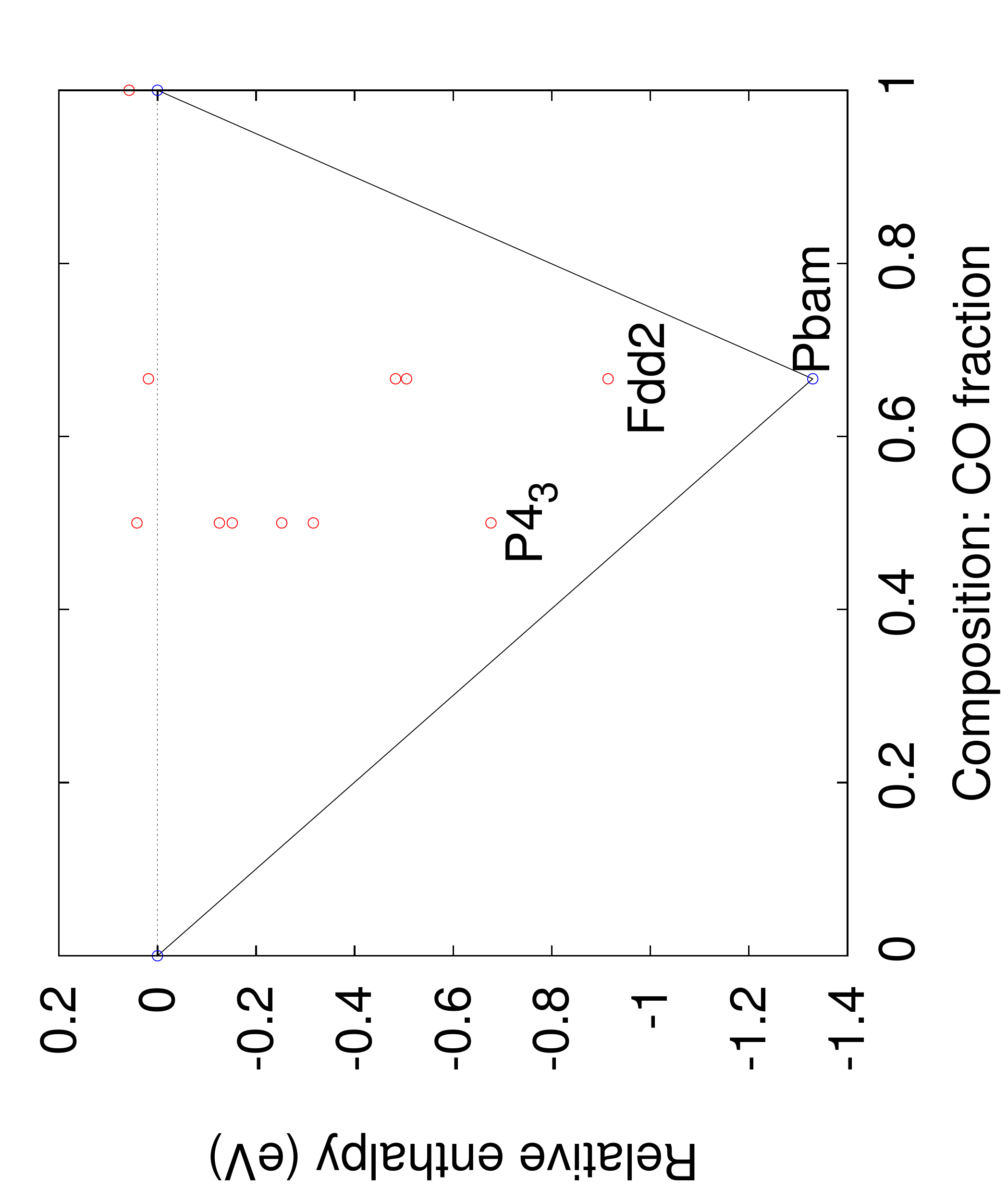}\end{turn}}
  \caption{Binary hulls for \ce{CO}/\ce{N_2}. Enthalpies relative to the most
  stable separate phases of carbon monoxide and nitrogen at the respective
  pressures. At \SI{20}{\giga\pascal}, these are molecular $P4_12_12$ N and
  three-dimensional framework $I2_12_12_1$ \ce{CO} structures, and at
  \SI{100}{\giga\pascal}, cg-N and layered $Cmcm$ \ce{CO}}
  \label{fig:hull}
\end{figure}

The binary hulls (Fig.~\ref{fig:hull}) show that the best structures for a
2$:$1 mixture of \ce{CO}$:$\ce{N_2} are more stable than the 1$:$1 mixture with
respect to the separate \ce{CO} and \ce{N_2} phases by up to
\SI{0.5}{\electronvolt}, and that the best structures found for a 1$:$1 mixture
are unstable because they are not on the convex hull. Searches using another
potentially interesting stoichiometry, \ce{CO_2}/\ce{N_2} (the carbon
dioxide/nitrogen system), yielded no stable structures.

\begin{figure}[ht]
  \subfloat[Polymeric $Pbam$ unit cell\label{fig:pPbam}]{\includegraphics[width=0.24\textwidth]{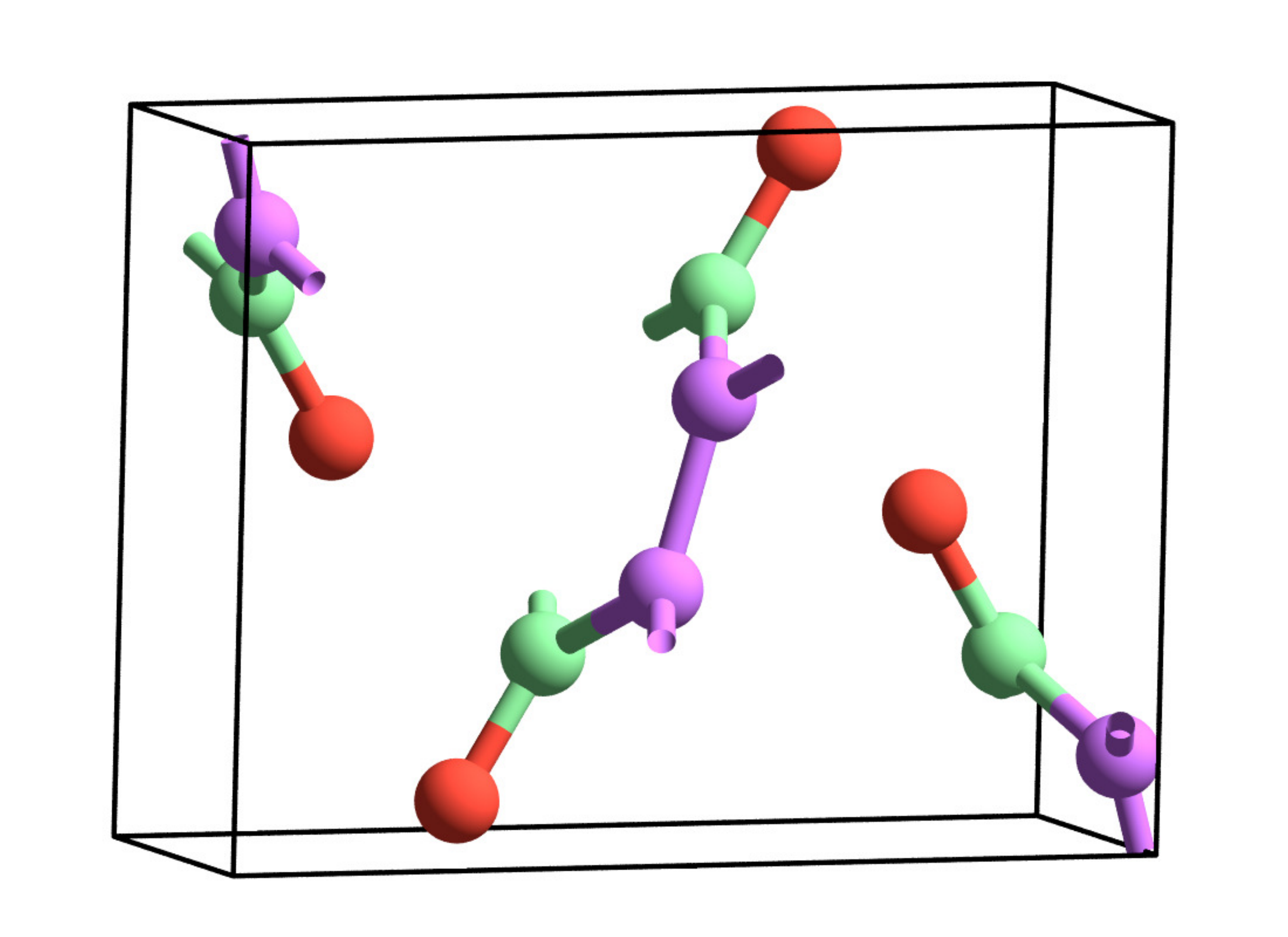}}
  \subfloat[Polymeric $Pbam$ stacking\label{fig:pPbam-stacking}]{\includegraphics[width=0.24\textwidth]{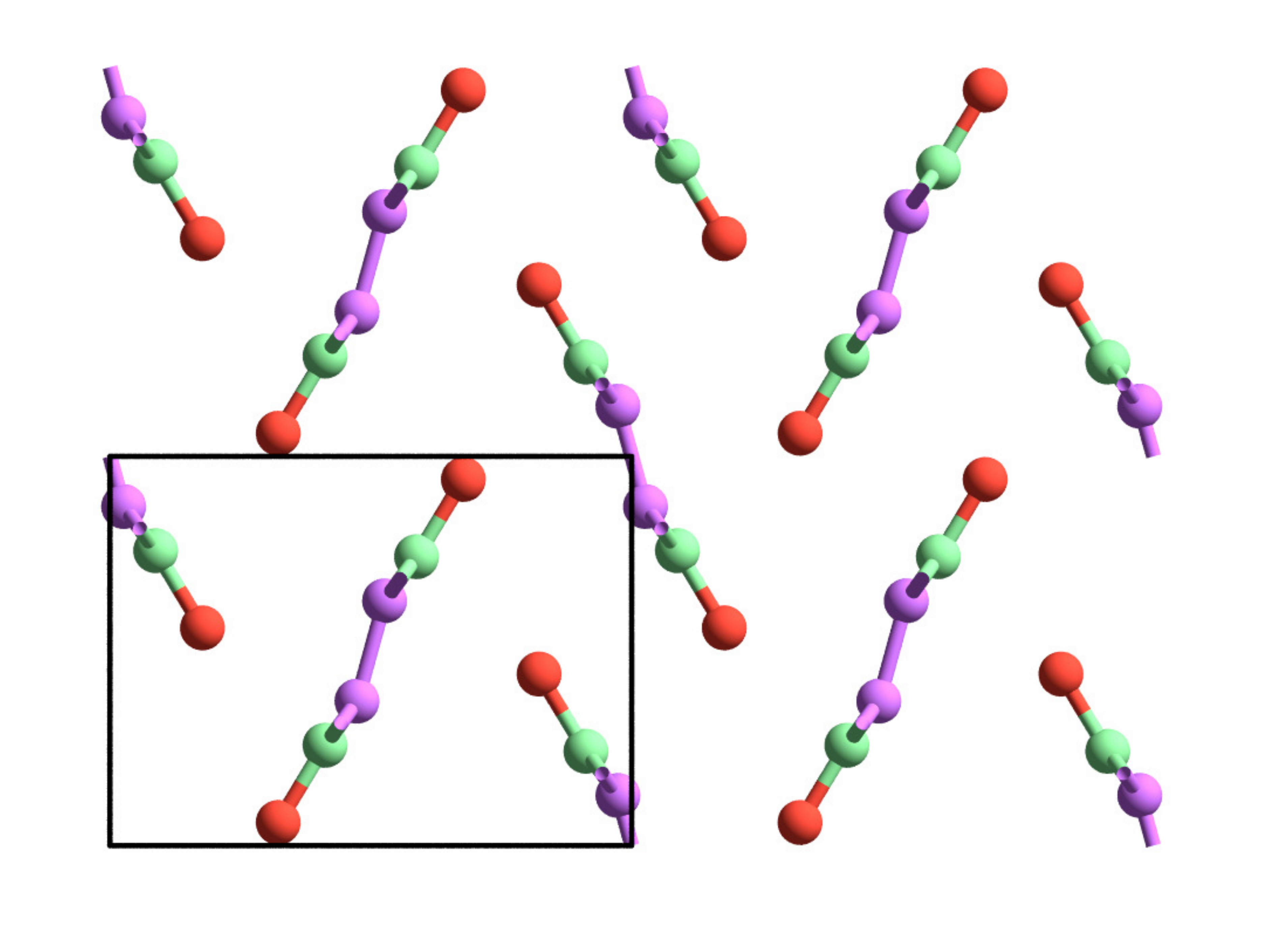}} \\
  \subfloat[$Fdd2$ unit cell\label{fig:Fdd2}]{\includegraphics[width=0.24\textwidth]{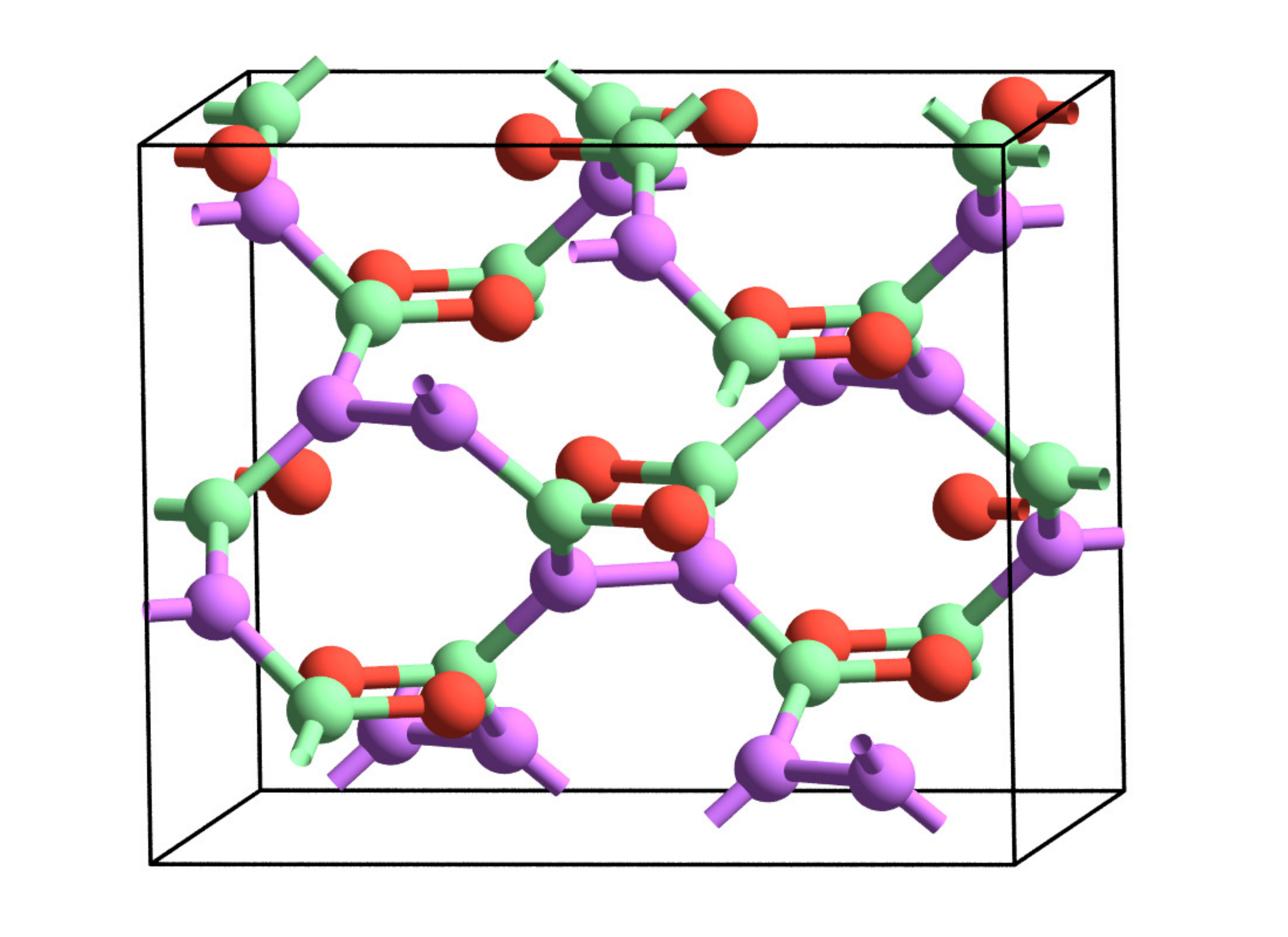}}
  \subfloat[$Fdd2$ stacking\label{fig:Fdd2-stacking}]{\includegraphics[width=0.24\textwidth]{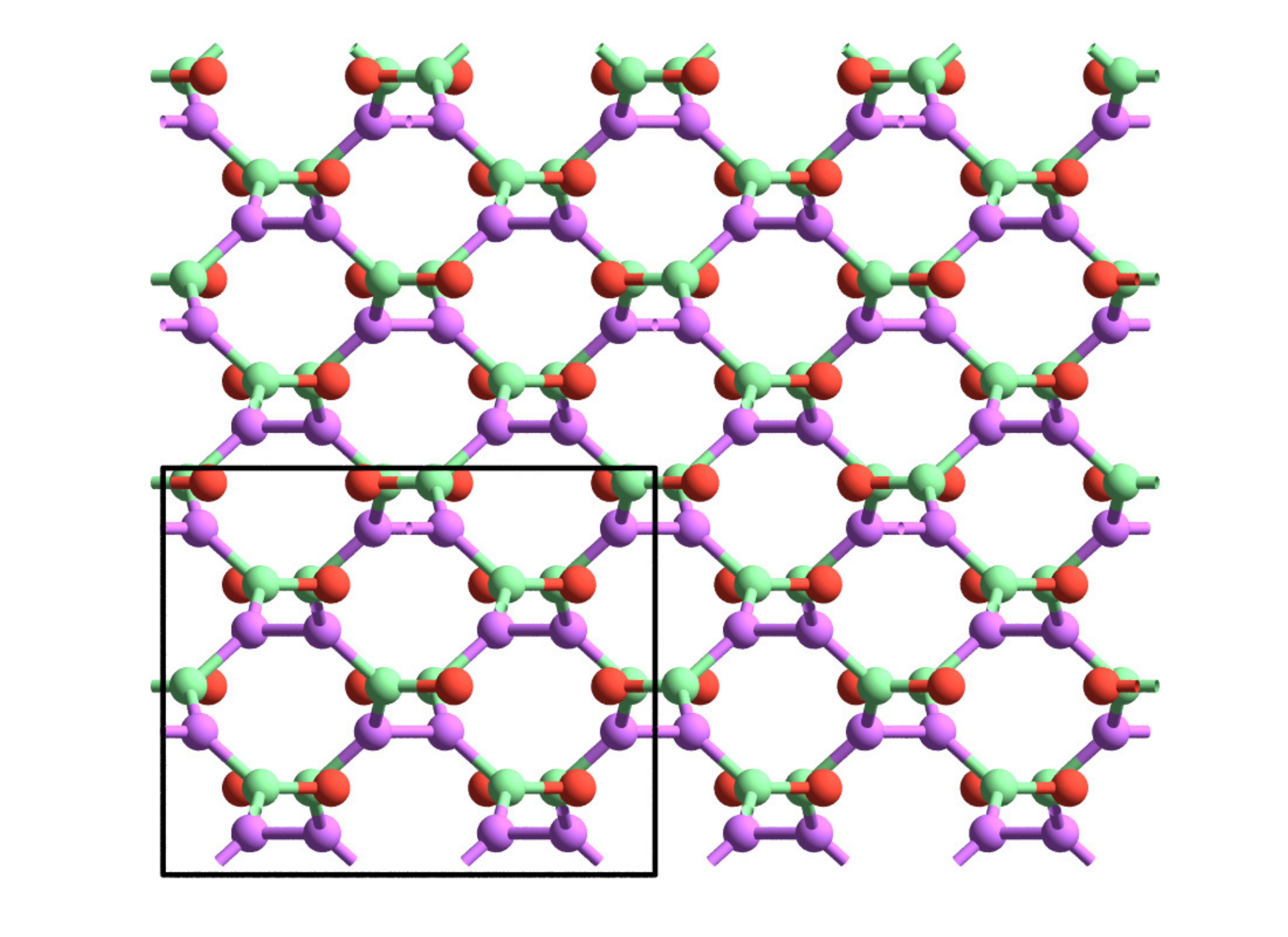}} \\
  \subfloat[Framework $Pbam$ unit cell\label{fig:Pbam}]{\includegraphics[width=0.22\textwidth]{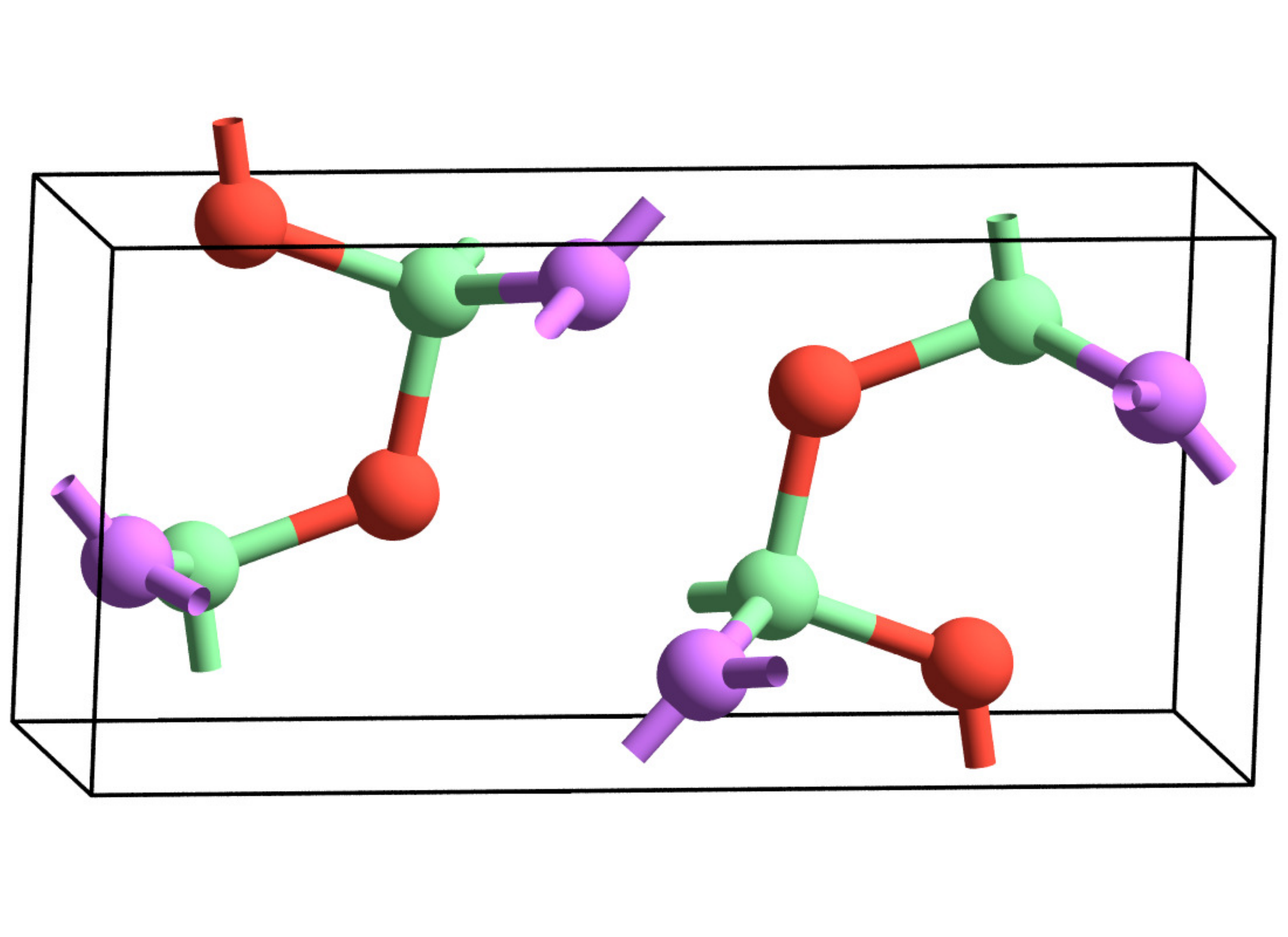}}\hspace{0.4cm}
  \subfloat[Framework $Pbam$ stacking\label{fig:3DPbam-stacking}]{\includegraphics[width=0.22\textwidth]{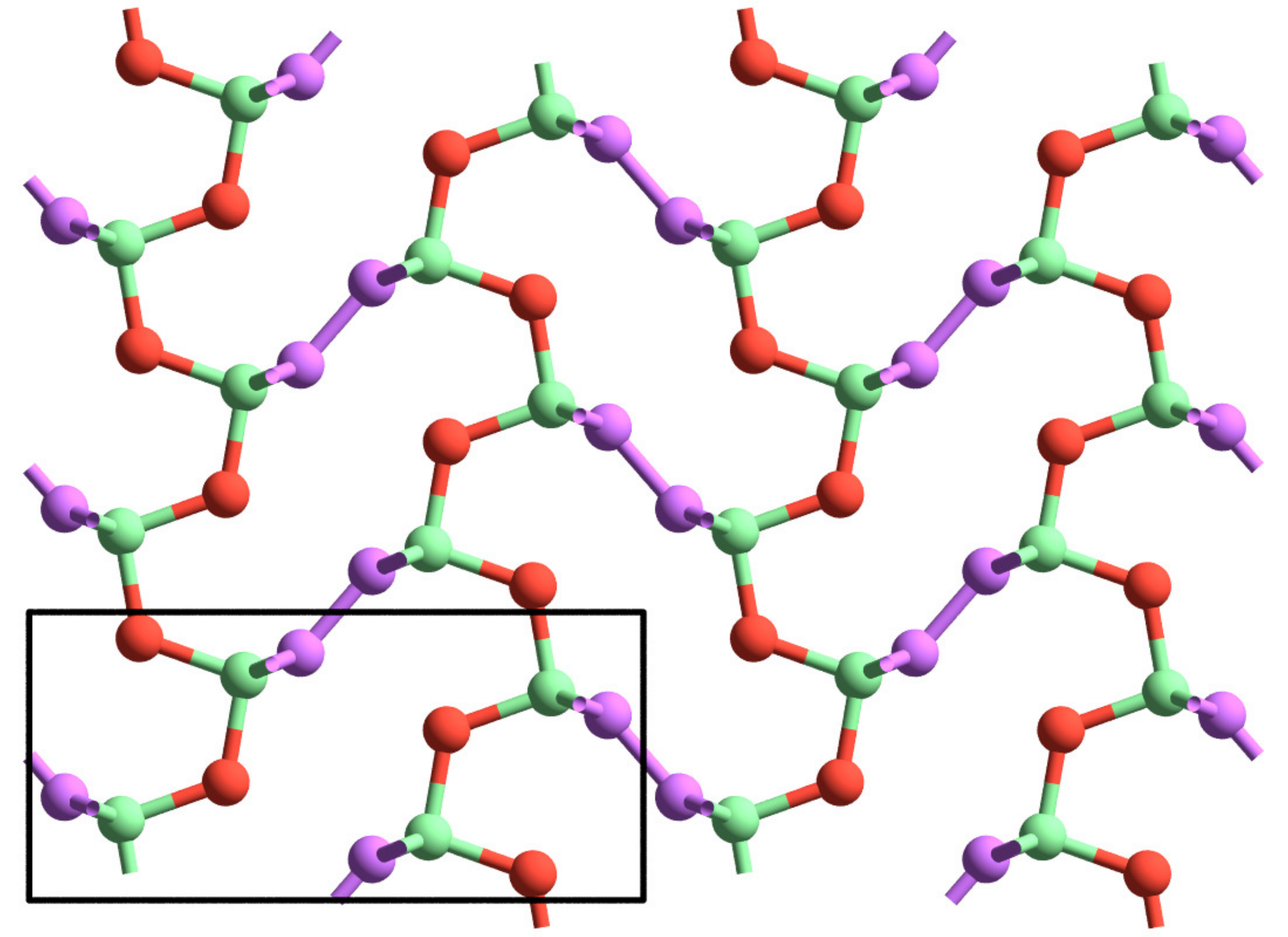}}
  \caption{Ground state structures for the CNO system (a 2$:$1 mixture of \ce{CO}$:$\ce{N_2}). Carbon atoms are green, nitrogen pink and oxygen red. Polymeric $Pbam$ is stable below \SI{18}{\giga\pascal}, $Fdd2$ between 18 and \SI{52}{\giga\pascal}, and framework $Pbam$ above \SI{52}{\giga\pascal}. The unit cells are on the left, and two-dimensional projections showing stacking on the right. Note the similarity in stacking between the two $Pbam$ phases in \ref{fig:pPbam-stacking} and \ref{fig:3DPbam-stacking}.}
\end{figure}

\begin{figure}[hb]
  \subfloat{\begin{turn}{-90}\includegraphics[width=0.22\textwidth]{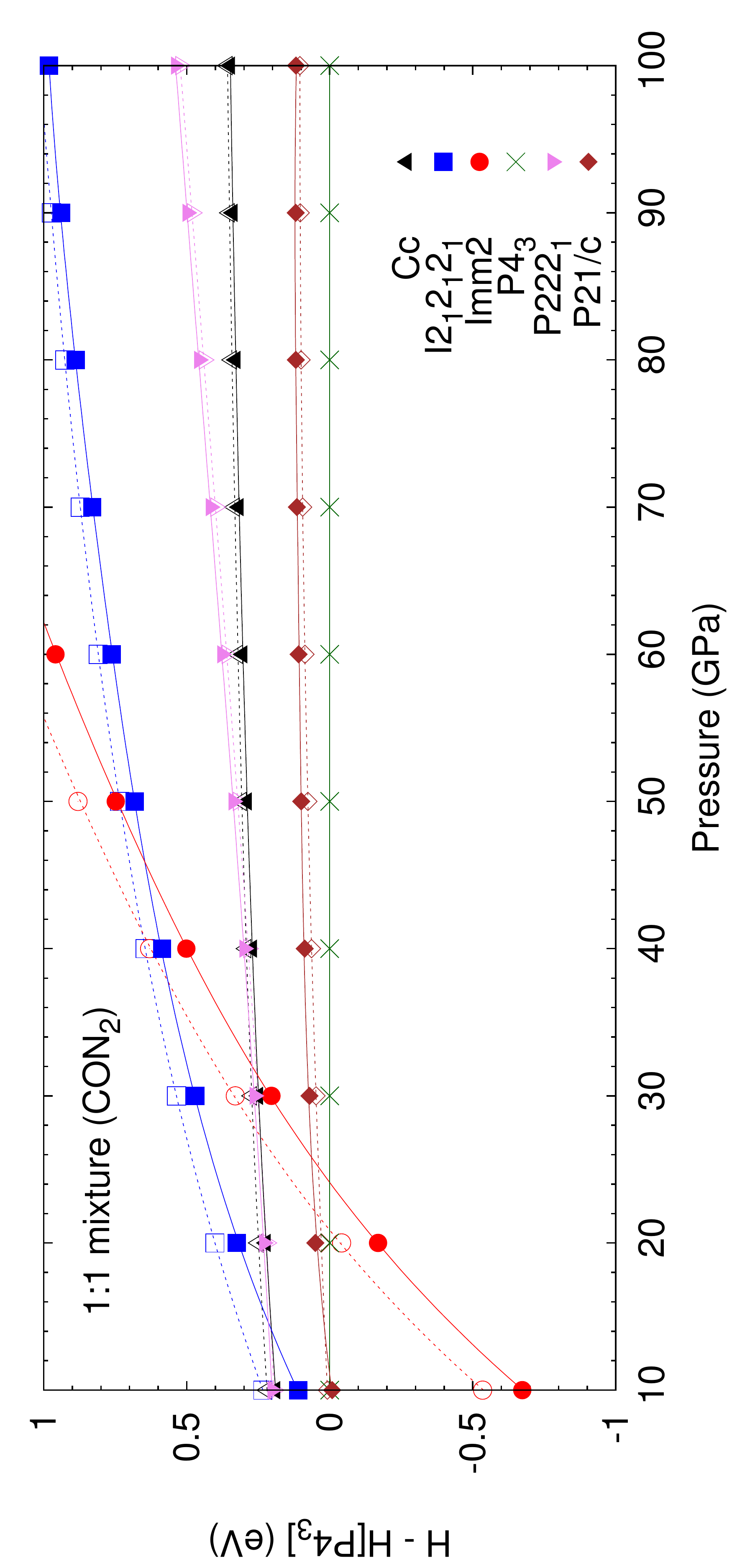}\end{turn}}\label{fig:CON2-eos} \\
  \subfloat{\begin{turn}{-90}\includegraphics[width=0.22\textwidth]{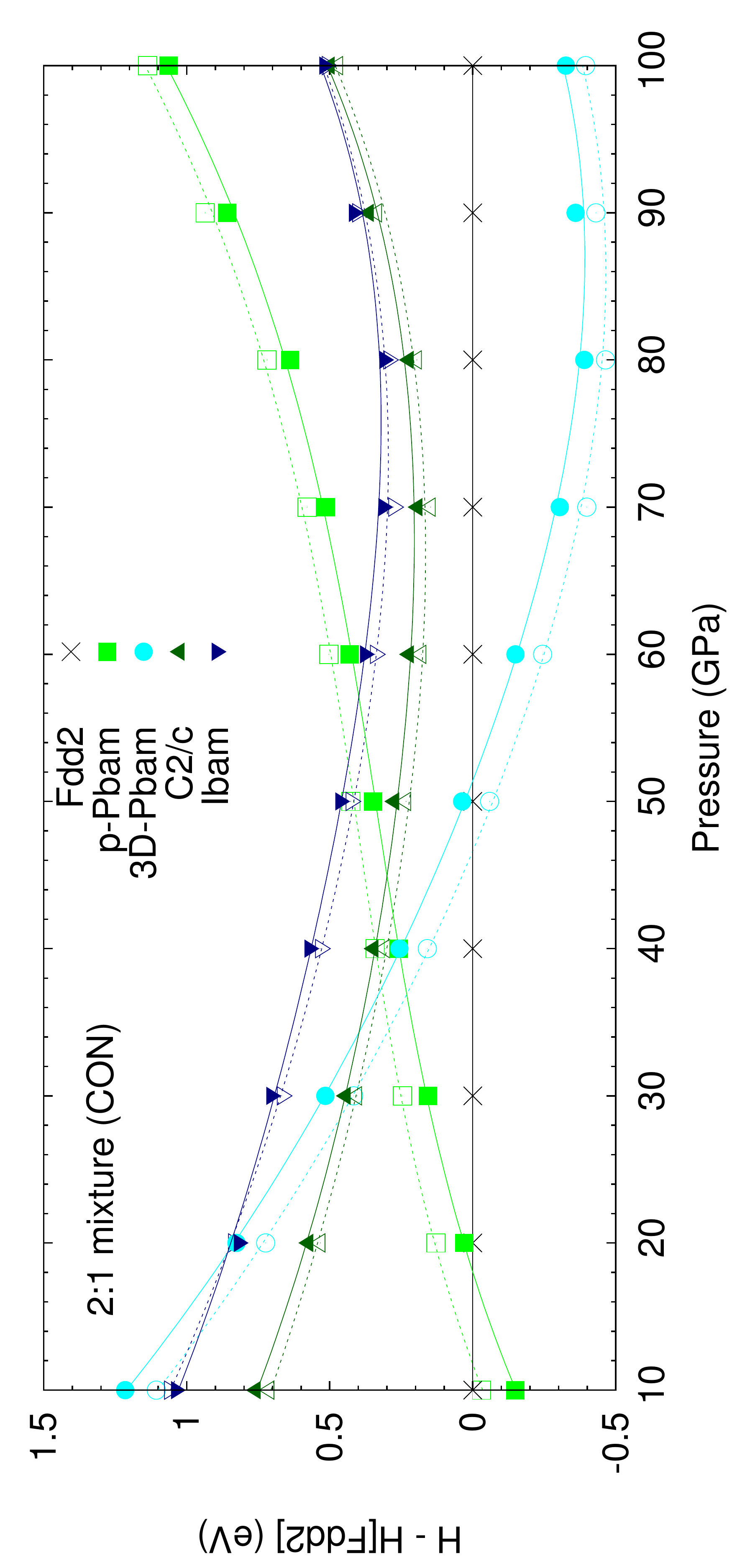}\end{turn}}\label{fig:CON-eos}
  \caption{Equation of state plots for \ce{CO}/\ce{N_2} mixture. Enthalpies are
  relative to the $P4_3$ phase for the 1$:$1 mixture, and relative to $Fdd2$
  for the 2$:$1 mixture. Broken lines represent calculations performed using a
  dispersion correction}
  \label{fig:eos}
\end{figure}

A search at \SI{20}{\giga\pascal} with the \ce{CNO} stoichiometry (i.e.\@ a
2$:$1 mixture) yielded the polymeric $Pbam$ structure (Fig.~\ref{fig:pPbam}).
This phase is the ground state at pressures up to \SI{18}{\giga\pascal}
according to Fig.~\ref{fig:CON-eos} and is more stable than its
\ce{CN_2O} counterpart ($Imm2$) because it does not exhibit signs of
phase separation.

The $Fdd2$ structure (Fig.~\ref{fig:Fdd2} is the most stable in the
\SIrange{18}{52}{\giga\pascal} range; notably, it contains \ce{C=O} double
bonds which degrade barrierlessly as the pressure increases beyond
\SI{50}{\giga\pascal}. As a result, it transforms into an energetically
unfavorable, unsaturated structure at pressures above \SI{52}{\giga\pascal},
and the $Pbam$ three-dimensional framework (Fig.~\ref{fig:Pbam}) becomes the
ground state.  Although the $Fdd2$ structure is the most stable for a
significant proportion of the pressure range, we have focussed on the phase
transition between the two $Pbam$ structures since they are much closer in
configuration space, and are likely to be separated by a much smaller kinetic
barrier. We expect the transition between the $Pbam$ structures to occur at
much lower temperatures than any transition involving the $Fdd2$ structure.

The three-dimensional framework $Pbam$ phase consists of near-planar
three-coordinated nitrogens, tetrahedral four-coordinated carbons and
two-coordinated oxygens, and is a large band gap insulator throughout the
pressure regime. The PBE band gap is \SI{2.95}{\electronvolt} at ambient
pressure, and \SI{3.35}{\electronvolt} at \SI{100}{\giga\pascal}, although it
should be noted that DFT consistently underestimates band gaps. Phonon
calculations performed at pressures in the range
\SIrange{0.1}{100}{\giga\pascal} yield no imaginary frequencies, suggesting
that it is dynamically stable throughout the pressure range, and significantly,
that the three-dimenstional framework phase can be recovered to ambient
pressure. The polymeric $Pbam$ structure is only dynamically stable above
approximately \SI{30}{\giga\pascal}. At \SI{20}{\giga\pascal} and below, it has
negative frequencies corresponding to an antiparallel motion of the weakly
bound polymer chains. Since it is conceivable that van der Waals interactions
may be significant for unsaturated polymeric structures at the low end of the
pressure scale, the pressure-enthalpy relationships were recomputed using a
semiempirical dispersion correction\citep{Grimme2006}. Figure \ref{fig:eos}
demonstrates that this correction results in a uniform shift in energy that is
approximately invariant with respect to pressure; however, the phase transition
pressures are reduced by \SIrange{5}{8}{\giga\pascal}.

The stability of framework $Pbam$ was tested against decomposition into likely
combinations of products from the ternary hull of the C-N-O system at
\SI{60}{\giga\pascal}. It was found to exothermically decompose to
$\alpha$-\ce{C_3N_4}, $I\bar{4}2d$ \ce{CO_2} and cg-N.

At ambient pressure, the metastable framework $Pbam$ phase is expected to
decompose exothermically to the same products as solid carbon monoxide
(graphitic carbon and molecular \ce{CO_2}) and cg-N (molecular \ce{N_2}):
\begin{align*}
\ce{4CNO -> 2CO_2 + 2C + 2N_2 +} \SI{4.1}{\electronvolt}
\end{align*}
The \emph{chemical} energy released during this reaction is estimated to be
\SI{4.1}{\electronvolt} at the PBE level (including a
semiempirical dispersion correction\citep{Grimme2006}). This corresponds to an
energy density of approximately \SI{2.2}{\kilo\joule\per\gram}, which compares
favorably to values in the range \SIrange{1}{3}{\kilo\joule\per\gram} for
modern explosives such as TATB, RDX and HMX\citep{Evans2006}.

\begin{figure}[ht]
  \begin{turn}{-90}
    \includegraphics[scale=0.32]{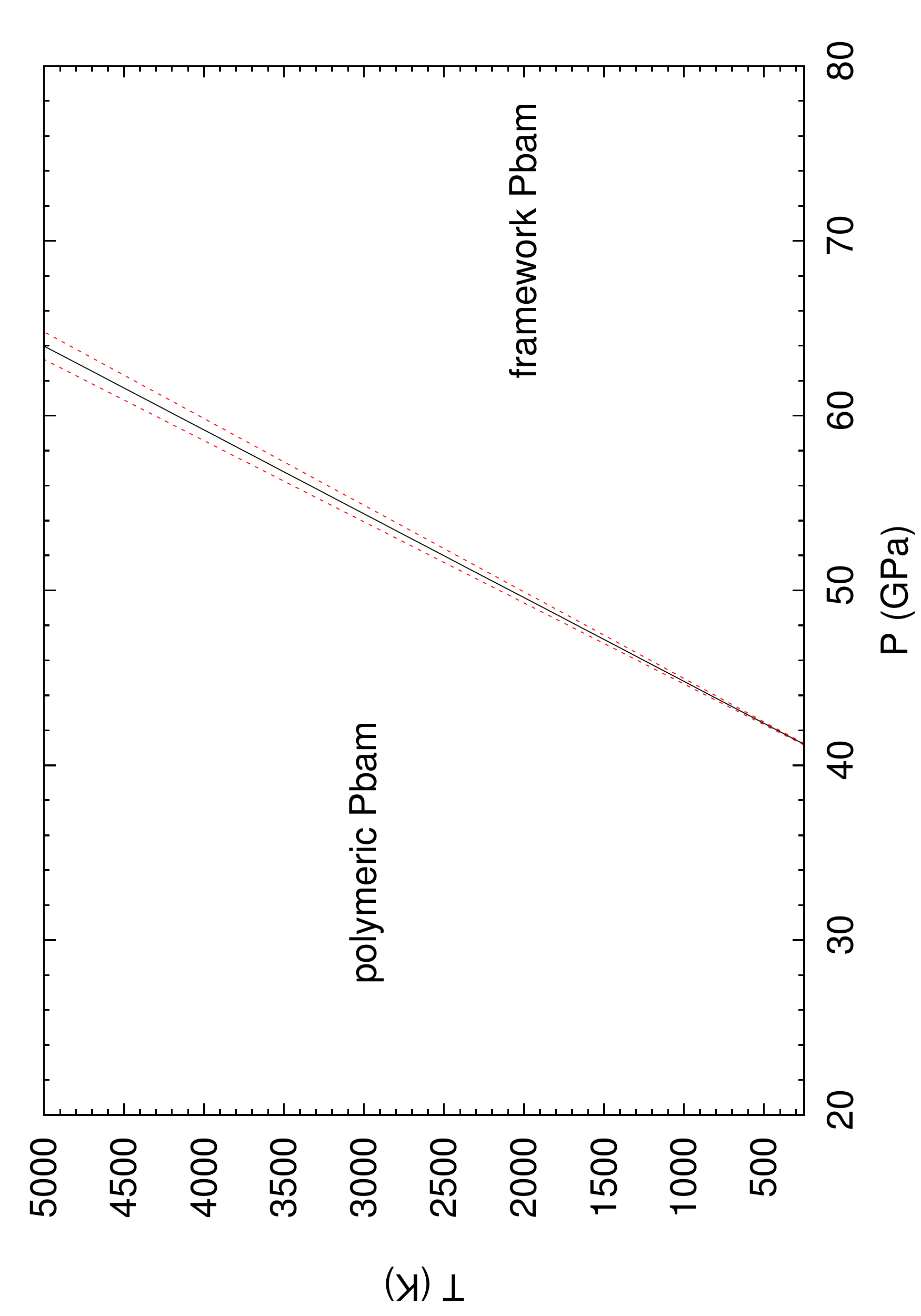}
  \end{turn}
  \caption{Theoretical pressure-temperature phase diagram. The line is
  calculated using the zero temperature phase transition pressure
  of \SI{40}{\giga\pascal} and a Clapeyron slope ($dP/dT = \Delta S/\Delta V$)
  of 4.80 $\pm$ \SI{0.16}{\mega\pascal\per\kelvin}. The broken lines denote the
  error in the gradient}
  \label{fig:pt}
\end{figure}

Both the polymeric and framework $Pbam$ structures have a six-ring motif
containing four 3-coordinated nitrogen and two carbon atoms that are sp$^2$
hybridized in the former (resulting in a \ce{C=O} double bond), and sp$^3$ in
the latter (two \ce{C-O} single bonds). Given the similarities in the stacking
of these units, a phase change is conceivable, although the separate lines in
Fig.~\ref{fig:CON-eos} suggest that there is an associated energy barrier.
This transition is expected to occur at \SI{40}{\giga\pascal} at zero
temperature.  By deriving the entropy from the phonon spectra, we estimated the
Clapeyron slope to obtain the temperature dependence of the transition pressure
(Fig.~\ref{fig:pt}).

The Knoop hardness of the framework $Pbam$ phase at ambient pressure was
calculated to be \SI{18}{\giga\pascal} using the electronegativity method
\citep{Li2008,Lyakhov2011}, which is below the superhardness threshold of
\SI{40}{\giga\pascal}. Its bulk modulus, \SI{288}{\giga\pascal}, is lower than
that of cubic gauche nitrogen which has been experimentally determined as over
\SI{300}{\giga\pascal}\citep{Eremets2004}. This, and the low Knoop hardness,
can be attributed to the large interstitial voids in this structure.

We find that a mixture of two parts \ce{CO} to one part \ce{N_2} polymerizes to
form a one-dimensional polymeric phase with $Pbam$ symmetry under undemanding
conditions compared with cg-N (below \SI{18}{\giga\pascal}). This low-pressure
transition is somewhat surprising considering the physical similarities of
\ce{CO} and \ce{N_2} molecules. This phase will transform to a
three-dimensional framework with the same symmetry at the relatively low
pressure of \SI{40}{\giga\pascal} at zero temperature. This framework
represents the best of both worlds; the carbon monoxide allows the mixture to
polymerize at a low pressure, and the nitrogen stabilizes the mixture in a
fully covalent crystalline form. It has a high energy density of
\SI{2.2}{\kilo\joule\per\gram} with respect to molecular nitrogen, carbon
dioxide and graphitic carbon, making it comparable to conventional explosives,
and it is dynamically stable at ambient pressure, fulfilling the crucial
requirements of a HEDM.

Z.R.~and A.M.S.~thank S.~Ninet, F.~Datchi and Y.~Le Godec for useful
discussions, and acknowledge the GENCI IDRIS and CINES French national
supercomputing facilities for CPU time (Project Grant No.~91387). Z.R. and A.M.S.
thank French Agence Nationale de la Recherche (ANR) for support through project
ANR-2011-BS08-018. C.J.P. is supported by the EPSRC. C.P. acknowledges funding
from the French Agence Nationale de la Recherche (ANR, Project No. 11-JS56-001
``CrIMin'').

\bibliographystyle{unsrt}
\bibliography{literature}

\end{document}